\documentclass[preprint,showpacs,preprintnumbers,amsmath,amssymb]{revtex4}
\usepackage{graphicx}

\begin{document}

\title{Enhancing the Conductance of a Two-Electron Nanomechanical Oscillator}
\author{ J. R. Johansson$^1$,  L. G. Mourokh$^{1,2,3}$, A. Yu. Smirnov$^{1,4}$, and Franco Nori$^{1,5}$}

\affiliation{ $^1$ Frontier Research System, The Institute of
Physical and
Chemical Research (RIKEN), Wako-shi, Saitama, 351-0198, Japan \\
$^2$ Department of Physics, Queens College, The City University of
New York, Flushing, New York 11367, USA \\ $^3$ Department of
Engineering Science and Physics, College of Staten Island, The
City University of
New York, Staten Island, New York 10314, USA \\
$^4$ CREST, Japan Science and Technology Agency, Kawaguchi,
Saitama, 332-0012, Japan \\
$^5$ Center for Theoretical Physics, Applied Physics Program, Center
for the Study of Complex Systems, Physics Department, The University
of Michigan, Ann Arbor, MI 48109-1040, USA}

\begin{abstract}
{We consider electron transport through a {\it mobile\/} island
(i.e., a nanomechanical oscillator) which can accommodate one or
two excess electrons and show that, in contrast to  {\it
immobile\/} islands, the Coulomb blockade peaks, associated with
the {\it first\/} and {\it second\/} electrons entering the
island, have different functional dependence on the
nano-oscillator parameters when the island coupling to its leads
is asymmetric. In particular, the conductance for the second
electron (i.e., when the island is already charged) is greatly
enhanced in comparison to the conductance of the first electron in
the presence of an external electric field. We also analyze the
temperature dependence of the two conduction peaks and show that
these exhibit different functional behaviors.}
\end{abstract}


\maketitle

\section{Introduction}

Electron transport in nanoelectromechanical systems (NEMS), such as
suspended nano-beams, cantilevers, and nano-oscillators, is now
attracting considerable attention \cite{reviews}. In shuttles,
electrons can be carried by a single nano-particle or single
molecule, which oscillates between two leads. This mechanical motion
strongly modifies the lead-shuttle tunneling matrix elements,
affecting the charge transfer. Theoretical
\cite{theory,McC1,DMLmech} and experimental \cite{Park1,exp} studies
of nanooscillators clearly demonstrated the influence of mechanical
motion on their electrical properties.

Previously, electron transport through a moving island was examined
in the strong Coulomb-blockade regime, when the conducting level of
the nano-oscillator can only be {\it single\/} populated, with
higher-energy states being energetically inaccessible. Here, we
demonstrate that a {\it charged\/} island behaves differently from
an uncharged one; correspondingly, the possible {\it double\/}
occupation of the conducting level leads to a situation where the
Coulomb blockade peaks, associated with the first and second
electrons transferred through the island, have {\it different\/}
dependencies on the nano-oscillator parameters. Moreover, we show
that the double occupation leads to a {\it conductance
enhancement\/} for the second electron entering the island. To
achieve that, we apply a previously-developed approach
\cite{Smirnov1,Smirnov2} which makes it possible to examine the case
of finite on-site Coulomb interaction.

It should be noted that the moving island studied in this work can
be considered as a {\it shuttle} because of its actual function:
shuttling. When the island moves closer to the left lead, increasing
that matrix element, it loads an electron. Then, the island moves
closer to the right lead and unloads the electron. Therefore, this
describes the operation of an electron shuttle. We have opted to use
the term ``mobile island'' for this ``electron shuttle'' because in
the nanomechanical community oscillators exhibiting an instability
are called ``shuttles''. Still, functionally, the moving island
described here is effectively a shuttle.

Usually, a nano-oscillator is considered to be placed
symmetrically between the leads. However, recently several works
discussed the situation when there is an asymmetry in the
lead-oscillator coupling produced either by the difference in the
tunnel matrix elements \cite{Flensberg,Fazio} or by the spatial
shift of the equilibrium oscillator position \cite{Milburn1}. In
the latter case, it was theoretically proposed that, if the island
is closer to one lead than to the other, the current through the
structure depends {\it exponentially} on this spatial shift (with
the tunnelling length $\lambda$), because the overlap integral of
the electron wave functions in the island and in the leads,
involved in the tunnel matrix elements, exponentially decreases
with distance. In the model of Ref.~\cite{Milburn1}, this small
displacement, shifting the island close to one of the leads, was
produced by the large magnetic field gradient acting on the spin
of the nitrogen or phosphorus impurity incorporated into their
model of a $C_{60}$ shuttle. Here we show that such displacement
can be achieved {\it naturally} in the island without impurities
with an excess electron in an electric field produced by the
source-drain voltage or an external capacitor (see Fig.~1).
 Moreover, this kind of spatial asymmetry can be associated with the
Jahn-Teller effect: when an orbital state of an ion is degenerate
for symmetry reasons, the ligands will experience forces driving the
system to a lower-symmetry configuration, lowering its energy.
Consequently, the ligand position between the two ions is {\it not
symmetric\/} and changes with the electron transfer from one ion to
the other. {\it Oscillations} of such ligands, either as oxygen
atoms in manganites \cite{Satpathy} or rare-earth atoms in filled
skutterudites \cite{Hotta}, were analyzed jointly with the
Jahn-Teller effect. However, the tunnelling length was assumed to be
{\it infinite\/} in Refs.~\cite{Satpathy,Hotta} and the dependence
of the tunnel matrix elements on the oscillator position was {\it
not\/} taken into account. Here we consider these effects and find a
remarkably rich behavior of the conductance of nano-oscillators, if
the matrix elements have an asymmetry as in
Refs.~\cite{Flensberg,Fazio}.

The present paper is organized as follows. Section II introduces the
pertinent Hamiltonian including all interactions. The equations of
motion for the electron creation/annihilation operators are derived
in Sec.~III. The equations for the electron populations and the
populations correlator are derived and solved in Sec.~IV. In Sec.~V
we obtain explicit expressions for the lead-to-lead current and
discuss the dependence of the conductance on the system parameters.
The conclusions of this work are presented in Sec.~VII.

\section{Formulation}

To examine electron transport through a moving island, we assume
that the island has a single spatial state which can be populated by
two electrons having opposing spin projections, $\sigma$ and
$\bar\sigma$, with {\it finite\/} on-site Coulomb interaction ($U_0
\neq \infty$). It should be noted that here we consider the
situation where the coupling of the nano-oscillator to the leads is
weak, so the Kondo-like correlations are not important. The
Hamiltonian of this system is given by ($\alpha =L,R$ for left,
right; $\sigma =1,2$ for spin up, down; $\bar\sigma = 2,1$ )
\begin{equation}
H = \sum_{\sigma} E_{\sigma} a_{\sigma}^+a_{\sigma} + U_0
a_{\sigma}^+\, a_{\sigma}\, a_{\bar\sigma}^+\, a_{\bar\sigma}  +
\sum_{k\alpha\sigma} E_{k\alpha\sigma}\, c_{k\alpha\sigma}^+\,
c_{k\alpha\sigma} + H_{{\rm osc}} + H_{{\rm tun}},
\end{equation}
where $a^+_{\sigma}$ ($a_{\sigma}$) are the creation
(annihilation) operators for the electrons in the island and
$c^+_{k\alpha\sigma}$ ($c_{k\alpha\sigma}$) are the creation
(annihilation) operators with wavevector $k$ in the $\alpha$-lead.
The tunneling term,
\begin{equation}
H_{{\rm tun}} = - \sum_{k\alpha\sigma} T_{k\alpha}\,
w_{\alpha}(x)\, c_{k\alpha\sigma}^+\, a_{\sigma}\, + H.\, c.,
\end{equation}
has tunneling amplitudes depending explicitly on the position $x$ of
the island as $$w_{\alpha}(x)=\exp \left( \frac{x}{\lambda_{\alpha}}
\right)$$ with the tunneling lengths $\lambda_L = - \lambda$ and $
\lambda_R = \lambda$ for the left and right leads, respectively. The
Hamiltonian of the nanomechanical oscillator also contains the
interaction between the charge stored in the oscillator and an
effective electric field $\cal{E}$, as
\begin{equation}
H_{{\rm osc}} = {p^2\over 2M} + {M\omega_0^2\, x^2\over 2} - e\,
{\cal E} x\sum_{\sigma} N_{\sigma}.
\end{equation}
This field ${\cal E}$ can be produced by the voltage applied to the
leads, by the Jahn-Teller effect, or even by an independently
controlled electric field, if the structure is placed inside an
external capacitor. Here, $N_{\sigma} = a_{\sigma}^+\, a_{\sigma}$
is the electron population operator, and $M$ and $\omega_0$ are the
effective mass and the resonant frequency of the nano-oscillator,
respectively.

After the unitary transformation $U = \exp \left\{ -ip\,
\sum_{\sigma} x_{\cal{E}}\, N_{\sigma}\right\}$, where
$x_{\cal{E}} = e\cal{E}$/$(M\omega_0^2)$, we obtain the usual
expression for the oscillator Hamiltonian, $H_{{\rm osc}} = p^2/2M
+ M\omega_0^2x^2/2$, and the modified electron operators,
\begin{equation}
a_{\sigma}' = U^+a_{\sigma}U = e^{-ipx_{\sigma}} a_{\sigma},
\end{equation}
and tunnel matrix elements,
\begin{equation}
w_{\alpha}'(x) = w_{\alpha}(x + \sum_{\sigma'}
x_{\cal{E}}N_{\sigma'} ).
\end{equation}
 Using the properties of Fermi operators
($N_{\sigma}^2 = N_{\sigma},\ a_{\sigma}N_{\sigma} = a_{\sigma},\
N_{\sigma}a_{\sigma}=0,\ ...$), we obtain
\begin{eqnarray}
w_{\alpha}(x + x_{\cal{E}}(N_{\sigma}+N_{\bar\sigma}) ) &=&
w_{\alpha\sigma}(x)  + [w_{\alpha}(x+x_{\cal{E}}) - w_{\alpha}(x)
] N_{\sigma} \nonumber \\ + [w_{\alpha}(x+x_{\cal{E}}) -
w_{\alpha}(x)] N_{\bar\sigma}
 &+&
 [w_{\alpha}(x+2x_{\cal{E}}) - 2w_{\alpha}(x+x_{\cal{E}}) + w_{\alpha}(x)] N_{\sigma} N_{\bar\sigma},
\end{eqnarray}
and
\begin{equation}
a_{\sigma} w_{\alpha}(x + x_{\cal{E}} (N_{\sigma}+N_{\bar\sigma}))
= a_{\sigma} w_{\alpha}(x + x_{\cal{E}}) + A_{\sigma} [
w_{\alpha}(x + 2x_{\cal{E}}) - w_{\alpha}(x + x_{\cal{E}}) ].
\end{equation}
Here, we have introduced the Fermi operator $A_{\sigma} =
N_{\bar{\sigma}}a_{\sigma}$. Accordingly, the tunneling term has a
form
\begin{equation}
H_{{\rm tun}} = - \sum_{k\alpha\sigma}T_{k\alpha}\,
c_{k\alpha\sigma}^+\, B_{\alpha\sigma} - H.\, c. ,
\end{equation}
where $B_{\alpha\sigma}$ is the Fermi operator given by
\begin{equation}
B_{\alpha\sigma} = a_{\sigma} (u_{\alpha} + v_{\alpha}\,
N_{\bar\sigma} ),
\end{equation}
with the first term responsible for the electron tunneling from
the unoccupied island and the second one describing the electron
transfer through the double-populated level. Here,
\begin{equation}
u_{\alpha} = u_{\alpha}^{mn}\, \rho_{mn}\ \ \ {\rm and} \ \ \
v_{\alpha} = v_{\alpha}^{mn}\, \rho_{mn},
\end{equation}
where
\begin{equation}
  \rho_{mn} = |m\rangle\langle
n| \ \ \  (m,n=0,1,..)
\end{equation}
are the eigenstates of the mechanical oscillator Hamiltonian, and
the matrix elements of the tunneling amplitudes are given by
\begin{equation}
u_{\alpha}^{mn} = \langle m |\
\exp\left(\frac{x}{\lambda_{\alpha}}\right)
 \exp(-ipx_{\cal{E}})\ |n\rangle
\end{equation}
 and
\begin{equation}
 v_{\alpha}^{mn} = u_{\alpha}^{mn}\,
\left[\exp\left(\frac{x}{\lambda_{\alpha}}\right) - 1\right].
\end{equation}

Equations (12,13) can be considered as a generalization of the
Frank-Condon factors \cite{FC} accounting for the overlap integral
of the vibrational states before and after the transition. It is
evident that the Frank-Condon factors are different for the first
and second electrons entering the island because the center of the
oscillations is shifted in the case of the charged island (see
Fig.~1). It should be emphasized that by introducing the operators
$B_{\alpha\sigma}$ we are able to derive the equations of motion
{\it analytically}, without the use of the Hartree-Fock
approximation, assuming only a {\it weak\,} lead-island tunnelling
coupling. From a general point of view, the method presented here is
equivalent to the master equation approach.

\section{Equations of Motion}

Equations of motion for the island electron operators obtained
from the Hamiltonian, Eq.~(1) are given by
\begin{eqnarray}
i\dot{a}_{\sigma} = E_{\sigma} a_{\sigma} + U_0 A_{\sigma} - \sum_{k\alpha\sigma} T_{k\alpha\sigma}^*(u_{\alpha\sigma}^+ + v_{\alpha\sigma}^+ N_{\bar{\sigma}} ) c_{k\alpha\sigma} + \nonumber\\
\sum_{k\alpha\sigma} T_{k\alpha\bar{\sigma}}
c_{k\alpha\bar{\sigma}}^+ a_{\sigma}a_{\bar{\sigma}}
v_{\alpha\bar{\sigma}} + \sum_{k\alpha\sigma}
T_{k\alpha\bar{\sigma}}^*  v_{\alpha\sigma}^+ a_{\bar{\sigma}}^+
 a_{\sigma} c_{k\alpha\bar{\sigma}}
 \end{eqnarray}
 and
 \begin{eqnarray}
i\dot{A}_{\sigma} = ( E_{\sigma} +U_0) a_{\sigma}
- \sum_{k\alpha\sigma} T_{k\alpha\sigma}^*(u_{\alpha\sigma}^+ + v_{\alpha\sigma}^+ ) N_{\bar{\sigma}}  c_{k\alpha\sigma} + \nonumber\\
\sum_{k\alpha\sigma} T_{k\alpha\bar{\sigma}}
c_{k\alpha\bar{\sigma}}^+ a_{\bar{\sigma}}a_{\sigma}
u_{\alpha\bar{\sigma}} + \sum_{k\alpha\sigma}
T_{k\alpha\bar{\sigma}}^* (u_{\alpha\bar{\sigma}}^+ +
v_{\alpha\bar{\sigma}}^+ )a_{\bar{\sigma}}^+
 a_{\sigma} c_{k\alpha\bar{\sigma}}.
 \end{eqnarray}
Accordingly, equations for the {\it ensemble averaged\,} island
populations can be written as
\begin{equation}
 \frac{d \langle N_{\sigma}\rangle}{dt} = -i \sum_{k\alpha\sigma} T_{k\alpha\sigma} \langle c_{k\alpha\sigma}^+ B_{\alpha\sigma}\rangle  + H.\,c.
 \end{equation}
 and
 \begin{equation}
\frac{d \langle {N}_{\sigma}N_{\bar{\sigma}}\rangle }{dt}  = -i
\sum_{k\alpha\sigma} T_{k\alpha\sigma} \langle c_{k\alpha\sigma}^+
A_{\alpha\sigma}\rangle (\langle u_{\alpha\sigma}\rangle + \langle
v_{\alpha\sigma}\rangle )
 -i \sum_{k\alpha\sigma} T_{k\alpha\bar{\sigma}} \langle c_{k\alpha\bar{\sigma}}^+ A_{\alpha\bar{\sigma}}\rangle
 (\langle u_{\alpha\bar{\sigma}}\rangle + \langle v_{\alpha\bar{\sigma}}\rangle )  + H.\,c.
 \end{equation}
The equation of motion for the electron operators in the leads are
given by
\begin{equation}
i \dot{c}_{k\alpha\sigma} = E_{k\alpha\sigma} c_{k\alpha\sigma} -
T_{k\alpha\sigma} B_{\alpha\sigma}.
\end{equation}
In the case of weak lead-island tunnel coupling, the solution of
this equation can be represented as
\begin{equation}
c_{k\alpha\sigma}(t) = c_{k\alpha\sigma}^{(0)}(t) -
T_{k\alpha\sigma} \int dt_1\; g_{k\alpha\sigma}^r(t,t_1)\;
B_{\alpha\sigma}(t_1),
\end{equation}
where $c_{k\alpha\sigma}^{(0)}(t)$ is the unperturbed electron
operator and $g_{k\alpha\sigma}^r(t,t_1)$ is the retarded Green
function of the electrons in the leads, given by
\begin{equation}
g_{k\alpha\sigma}^r(t,t_1) = -i \langle
[c_{k\alpha\sigma}^{(0)}(t),c_{k\alpha\sigma}^{(0)+}(t_1)]_+
\rangle \theta(t-t_1)= - i e^{-iE_{k\alpha\sigma}(\tau )}
\theta(\tau ),
\end{equation}
where $[...,...]_+$ is the anticommutator, $\tau =t-t_1$, and
$\theta(\tau )$ is the unit step function. It should be emphasized
that the non-Markovian dynamics involved in Eq.~(19) allows us to
reveal manifestations of the oscillatory mechanical motion during
the tunneling events.

\section{Electron populations and populations correlator}

\subsection{Free-evolution approximation}

To determine electron populations in the island and the correlator
of the populations having different spin projections, we
substitute Eq.~(19) into Eqs.~(16,17). The correlators of the type
$\langle c_{k\alpha\sigma}^{(0)+}(t) B_{\alpha\sigma}(t) \rangle$
can be rewritten using the formula
\begin{equation}
\sum_{k\alpha\sigma} T_{k\alpha\sigma} \langle
c_{k\alpha\sigma}^{(0)+}(t)B_{\alpha\sigma}(t)\rangle = -
\sum_{k\alpha\sigma} |T_{k\alpha\sigma}|^2 \int^t_{-\infty} dt_1
\langle
c_{k\alpha\sigma}^{(0)+}(t)c_{k\alpha\sigma}^{(0)}(t_1)\rangle
\langle [ B_{\alpha\sigma}(t),B^+_{\alpha\sigma}(t_1)]_+ \rangle .
\end{equation}
To decouple the correlators for the electron operators in the
island, we use the approximation of their free evolution, which is
valid in the case of weak lead-island tunneling. The free
evolutions of the operators $a_{\sigma}, A_{\sigma},$ and
$B_{\alpha\sigma}$ are given by
\begin{equation}
a_{\sigma}(t) = e^{-iE_{\sigma}\tau } [ a_{\sigma} (t_1)  - ( 1 -
e^{-iU_0\tau} ) A_{\sigma}(t_1) ],
\end{equation}
\begin{equation}
 A_{\sigma}(t) =
e^{-i(E_{\sigma}+U_0)\tau} A_{\sigma}(t_1),
\end{equation}
and
\begin{equation}
 B_{\alpha\sigma}(t)  = e^{-iE_{\sigma}\tau } [a_{\sigma}(t_1) - A_{\sigma}(t_1)] u_{\alpha\sigma}(t) +
e^{-i(E_{\sigma}+U_0)\tau } A_{\sigma}(t_1) (
u_{\alpha\sigma}+v_{\alpha\sigma} )(t).
\end{equation}
Accordingly,
\begin{eqnarray}
\langle B_{\alpha\sigma}(t)B_{\alpha\sigma}^+(t_1) \rangle =
e^{-iE_{\sigma}\tau } \langle
u_{\alpha\sigma}(t)u_{\alpha\sigma}^+(t_1)\rangle \langle 1 -
N_{\sigma} - N_{\bar{\sigma}} +
N_{\sigma} N_{\bar{\sigma}} \rangle + \nonumber\\
e^{-i(E_{\sigma}+U_0)\tau } \langle
(u_{\alpha\sigma}(t)+v_{\alpha\sigma}(t))(u_{\alpha\sigma}^+(t_1)+v_{\alpha\sigma}^+(t_1))\rangle
\langle  N_{\bar{\sigma}} - N_{\sigma} N_{\bar{\sigma}} \rangle ,
\end{eqnarray}
and
\begin{eqnarray}
\langle B_{\alpha\sigma}^+(t_1)B_{\alpha\sigma}(t) \rangle =
e^{-iE_{\sigma}\tau } \langle
u_{\alpha\sigma}^+(t_1)u_{\alpha\sigma}(t)\rangle \langle
N_{\sigma} -
N_{\sigma} N_{\bar{\sigma}} \rangle + \nonumber\\
e^{-i(E_{\sigma}+U_0)\tau } \langle
(u_{\alpha\sigma}^+(t_1)+v_{\alpha\sigma}^+(t_1))(u_{\alpha\sigma}(t)+v_{\alpha\sigma}(t))\rangle
\langle  N_{\sigma} N_{\bar{\sigma}} \rangle .
\end{eqnarray}
The free-evolution approximation can also be used to calculate the
correlators of the mechanical operators. Using $\rho_{mn}(t) =
e^{i\omega_{mn}(t-t_1)} \rho_{mn}(t_1)$, we obtain:
\begin{eqnarray}
\langle u_{\alpha\sigma}(t)u_{\alpha\sigma}^+(t_1)\rangle =
\sum_{mn} |u_{\alpha\sigma}^{mn}|^2 e^{i\omega_{mn}\tau}
\langle \rho_m\rangle, \nonumber\\
\langle u_{\alpha\sigma}^+(t)u_{\alpha\sigma}(t_1)\rangle =
\sum_{mn} |u_{\alpha\sigma}^{mn}|^2 e^{-i\omega_{mn}\tau} \langle
\rho_n\rangle,
\end{eqnarray}
where $\langle \rho_n\rangle = \langle \rho_{nn}\rangle$ is the
steady-state distribution of the mechanical degrees of freedom and
$\omega_{mn} = \epsilon_m - \epsilon_n = \omega_0 (m-n)$.

\subsection{Electron occupations}

In the absence of an external magnetic field, the averaged electron
populations, $\langle N_1\rangle$ and $\langle N_2\rangle$, should
be equal: $\langle N_1\rangle =\langle N_2\rangle = \langle
N\rangle.$ As a result, we obtain the following equations for the
averaged electron occupation $\langle N\rangle$ and for the
correlation function of the populations with opposite spin
projections, $\langle N_1 N_2\rangle$,
\begin{equation}
\eta_1 \langle N_1 N_2\rangle = \eta_2 \langle N \rangle, \ \ \eta_3 \langle N\rangle + \eta_4 \langle N_1 N_2\rangle = \eta_0,
\end{equation}
having the simple solutions
\begin{equation}
\langle N\rangle = \frac{\eta_0 \eta_1}{\eta_1 \eta_3 + \eta_2
\eta_4}, \ \  \langle N_1 N_2\rangle = \frac{\eta_0 \eta_2}{\eta_1
\eta_3 + \eta_2 \eta_4} .
\end{equation}
We introduce the following coefficients
\begin{eqnarray}
\eta_0 = \sum_{\alpha} \sum_{mn} \Gamma_{\alpha} |u_{\alpha}^{mn}|^2 \langle \rho_m\rangle f_{\alpha}(E_0-\omega_{mn}),
\nonumber\\
\eta_1 = \sum_{\alpha} \sum_{mn} \Gamma_{\alpha} |u_{\alpha}^{mn}|^2 e^{2x_{\cal{E}}/\lambda_{\alpha}}
[ \langle \rho_n\rangle + \langle \rho_m - \rho_n\rangle f_{\alpha}(E_0 +U_0-\omega_{mn}) ], \nonumber\\
\eta_2 = \sum_{\alpha} \sum_{mn} \Gamma_{\alpha} |u_{\alpha}^{mn}|^2 e^{2x_{\cal{E}}/\lambda_{\alpha}} \langle \rho_m \rangle
f_{\alpha}(E_0 +U_0-\omega_{mn}),
\nonumber\\
\eta_3 = \sum_{\alpha} \sum_{mn} \Gamma_{\alpha} |u_{\alpha}^{mn}|^2 [ \langle \rho_n\rangle + \langle \rho_m - \rho_n\rangle
f_{\alpha}(E_0-\omega_{mn}) + \nonumber\\ \langle \rho_m\rangle f_{\alpha}(E_0-\omega_{mn}) - e^{2x_{\cal{E}}/\lambda_{\alpha}}
\langle \rho_m
\rangle f_{\alpha}(E_0+U_0-\omega_{mn})], \nonumber\\
\eta_4 = \sum_{\alpha} \sum_{mn} \Gamma_{\alpha} |u_{\alpha}^{mn}|^2 [ \langle \rho_n - \rho_m \rangle f_{\alpha}(E_0 -
\omega_{mn}) - \nonumber\\ e^{2x_{\cal{E}}/\lambda_{\alpha}}\langle \rho_n - \rho_m \rangle f_{\alpha}(E_0 +U_0 - \omega_{mn}) +
(e^{2x_{\cal{E}}/\lambda_{\alpha}} - 1 ) \langle \rho_n\rangle ].
\end{eqnarray}
Here, $f_{\alpha}(E)$ are the electron Fermi distribution
functions in the corresponding lead and $\langle ... \rangle$
means ensemble averaging. In the wide-band limit, we can introduce
the tunnel rate as
\begin{equation}
\Gamma_{\alpha\sigma} = 2\pi \sum_k |T_{k\alpha\sigma}|^2
\delta(\omega - E_{k\alpha\sigma} ).
\end{equation}
In this work, we examine the case of a very small source-drain
voltage applied to the system, so the density matrix of the
mechanical oscillator has the equilibrium form $(k_B=1)$
\begin{equation}
\rho_m =  e^{-\hbar \omega_0 m /T} \left( 1 - e^{-\hbar \omega_0/T} \right).
\end{equation}

We plot the solutions, Eq.~(29), as well as the second cumulant,
\begin{equation}
K_N= \langle N_{1} N_{2} \rangle - \langle N\rangle^2,
\end{equation}
in Fig.~2 as functions of the separation between the energy of the
island level $E_{\sigma}$ and the equilibrium chemical potential
$\mu$ of the leads. The following set of parameters, associated with
$C_{60}$, was chosen \cite{Park1}: the charging energy,
$U_0=270$~meV, the fundamental frequency, $\hbar\omega_0=5$~meV, and
the fundamental uncertainty of the oscillator position
$$r_0=\sqrt{\hbar/2M\omega_0}=3.8 \ \textrm{pm}. $$ The magnetic field is taken to
be zero (so $E_{\sigma}=E_{\bar\sigma}$), $T=77$~K, $eV=0.5$~meV,
and $\lambda =4$~pm. It is evident from Fig.~2 that when the
electron energy level on the island becomes smaller than $\mu$
(modulo thermal broadening), the island is single-populated and,
when the energy separation between $E_{\sigma}$ and $\mu$ is larger
than the charging energy, the island is double-populated, as
expected. It should be emphasized that although the {\it ensemble
averaged} values of both electron populations are nonzero in the
case of single occupation, the population correlator is zero,
meaning that the electron having only one of the spin projections
can be found in the specific sample. This Pauli repulsion also
manifests itself in the negative value of the cumulant $K_N$ in the
single-occupation regime. It should be also noted that the
functional dependencies of Fig.~2 do {\it not} depend on the value
of $x_{\cal{E}}$ and the asymmetry of the couplings to the left and
right leads.

\section{Electron current and conductance}

The current flow of electrons having $\sigma$-projection of the
spin from the $\alpha$-lead can be defined as
\begin{equation}
I_{\alpha\sigma} = e\frac{d}{dt} \sum_k \langle
c_{k\alpha\sigma}^+ c_{k\alpha\sigma}\rangle  = ie \sum_k
T_{k\alpha\sigma}\langle
c_{k\alpha\sigma}^+B_{\alpha\sigma}\rangle + H.\,c.
\end{equation}
Using the same approximations as in the previous section, we
obtain
\begin{eqnarray}
I_{\alpha} = e \Gamma_{\alpha}\sum_{mn} |u_{\alpha}^{mn}|^2 \left\{ (1-f_{\alpha}(E_{0}-\omega_{mn})) \langle N - N_{1} N_{2}
\rangle \langle \rho_n\rangle \right. \nonumber \\ - f_{\alpha}(E_{0}-\omega_{mn})\langle 1 - 2 N_0 + N_{1} N_{2} \rangle \langle
\rho_m\rangle
\nonumber\\
+  e^{2x_{\cal{E}}/\lambda_{\alpha}} [(1-f_{\alpha}(E_{0} + U_0 -\omega_{mn}))\langle N_{1} N_{2} \rangle \langle \rho_n\rangle
\nonumber \\ \left. - f_{\alpha}(E_{0} + U_0 -\omega_{mn})\langle N - N_1 N_2 \rangle \langle \rho_m\rangle ]\right\} \  .
\end{eqnarray}
The associated conductance,
\begin{equation}
G = 2\ \frac{I_L}{V},
\end{equation}
 is presented in Fig.~3 as a
function of $(E_{0}-\mu)$ and $x_{\cal{E}}$, using the same
parameters as in Fig.~2 with coupling constants $h\Gamma_L$ = 0.1
meV and $h\Gamma_R$ = 0.002 meV and temperatures (a) $T=4$~K and (b)
$T=77$~K. The projections of the three-dimensional plots unto both
the ``$G$ versus $(E_{\sigma}-\mu)$" and ``$G$ versus $x_{\cal{E}}$"
planes are shown in Fig.~3(c). One can see from Fig.~3 that the
magnitudes of the conductance peaks, associated with the first and
second electrons entering the island, are only equal to each other
for $x_{\cal{E}}=0$ (conventional Coulomb blockade case). Moreover,
the conductance through the {\it charged} island is {\it drastically
enhanced} at positive moderate values of $x_{\cal{E}}$. It should be
noted that the electric field-induced shift would not produce a
conductance enhancement for the {\it immobile\,} island because the
exponential increase of the tunnel matrix element between the island
and one of the leads is compensated by the same exponential decrease
of the tunnel matrix element coupling to the other lead. However,
for the {\it mobile\,} island, these matrix elements are averaged
over the island oscillatory motion and the shift is not cancelled
out. This is even more pronounced for the charged island where the
center of the oscillations is already shifted by the presence of the
first electron. Formally, the account of the oscillatory mechanical
motion during the tunneling events becomes possible due to the {\it
non-Markovian\/} character of the equation of motion. The dependence
of the conductance peaks on $x_{\cal{E}}$ has a Gaussian form
(coming from the Frank-Condon factors) with the centers shifted to
two different positive values of $x_{\cal{E}}$. With increasing
temperature, the conductance peaks become broader and the shift is
increased, as seen in Fig.~3. This shift can be attributed to the
phonon-blockade effect discussed in Refs. \cite{Flensberg,FC}. It
should be noted that the bias-voltage independent displacement
$x_{\cal{E}}$ can be created, for example, by nearby charge
impurities, image charges, device geometry, etc. \cite{McC1}.

We also examine the temperature dependence of the conductance peak
magnitude. For the {\it immobile} island, one can expect either no
temperature dependence, in the case of quantum-mechanical tunneling,
or thermal-activation dependence, in the case of over-the-barrier
hopping transport. However, deviations from such behaviors were
observed both in transport through single molecules \cite{Ratner2}
and in the resistivity of manganites \cite{Noginova1}.
Theoretically, it was shown that either the mechanical motion of the
nanoconductor or coupling to the quantized thermal modes
\cite{Smirnov1,Lundin1} can lead to exotic types of temperature
dependence. These can be seen in Fig.~4 for various values of
$\lambda$, at $x_{\cal{E}}=0$. It should be noted that the curves
are identical for both peaks in this case. For nonzero
$x_{\cal{E}}$, the temperature dependence becomes even more
complicated for small $\lambda$, as can be seen in Fig.~5(a) and
5(b) for the charged and uncharged shuttle, respectively, because of
the temperature-induced shift of the peak position (see Fig.~3). It
should be emphasized that the functional dependencies at small
$\lambda$ are very {\it different\/} for the two conduction peaks,
with the peak for the second electron being almost an order of
magnitude larger.

The current $I_{\alpha\sigma}$ through the shuttle is extremely
sensitive to the value of $\lambda$, because it appears in several
exponents of Eq.~(35). It is evident from Fig.~4 that the smaller
$\lambda$ is, the larger the conductance of the system becomes.
Therefore, the quality of the leads plays a more important role in
the electrical properties of nano-oscillators than in most
standard electronic devices.

\section{Conclusions}

In conclusion, we have examined electron transport through a mobile
island which can contain one or two electrons. We have derived the
equations of motion for the electron creation/annihilation operators
and have been able to evaluate them ({\it without\/} using the
Hartree-Fock approximation) by introducing a complex Fermi operator
$B_{\alpha\sigma}$, Eq.~(9). Based on this microscopic approach, the
equations for the island populations and the correlator of the
populations have been derived and solved. They are involved in the
expression for the electron current through the structure, also
obtained microscopically. We have shown that in the presence of an
external electric field (produced either by the voltage applied to
the system, by the Jahn-Teller effect in the molecular junctions, or
by an external capacitor), and an asymmetry in the coupling of the
island to the leads, the conductance of the second electron entering
the {\it charged\/} island is greatly enhanced. The temperature
dependence of the conductance has been also discussed.

This work was supported in part by the National Security Agency,
Laboratory of Physical Sciences, Army Research Office, JSPS CTC
Program, and National Science Foundation grant No. EIA-0130383.

\newpage

\begin{figure}[ht]
\includegraphics[width=8.0cm]{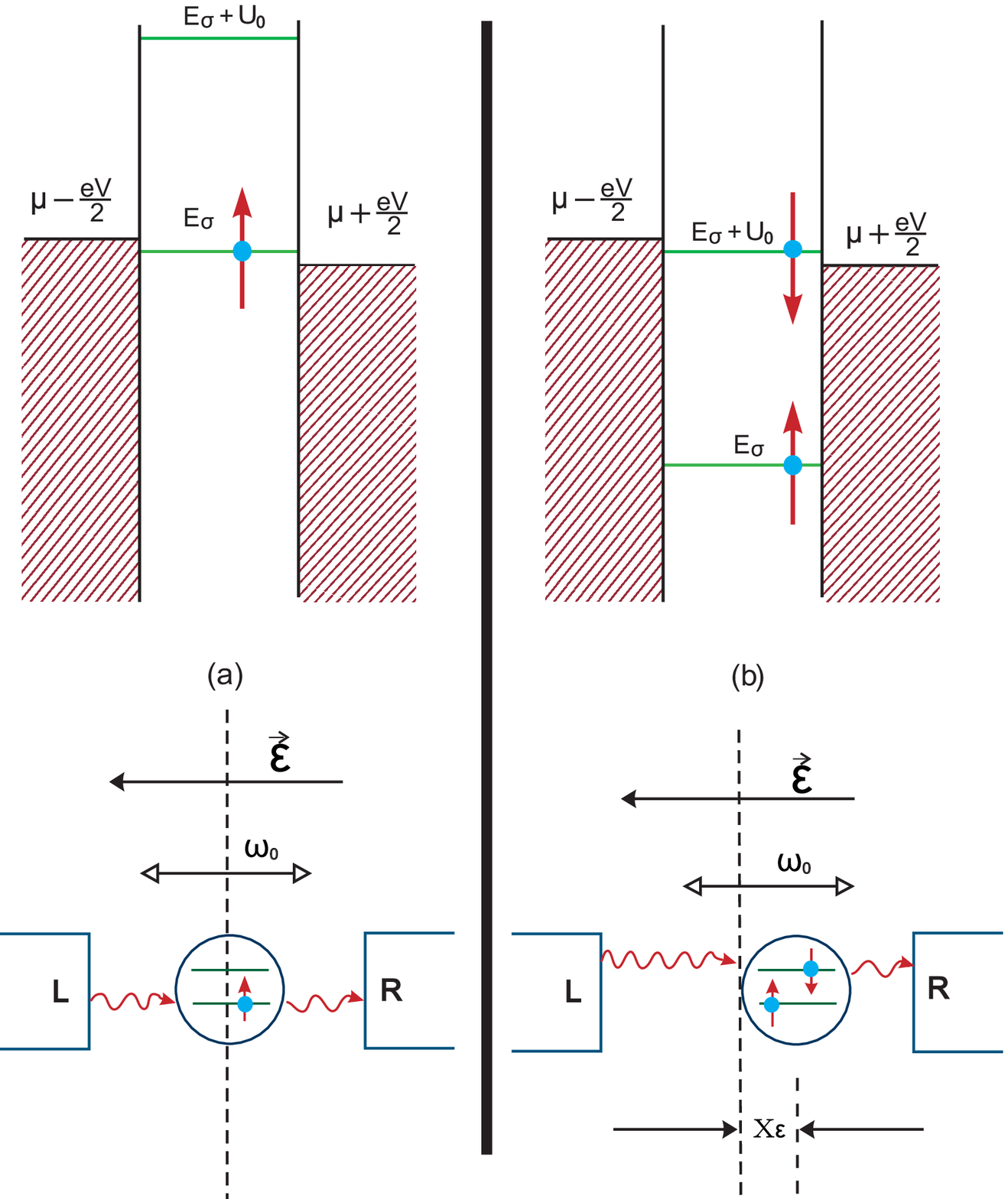}
\caption{\label{fig1} (Color online) Schematic diagram of electron
transport through (a) uncharged and (b) charged nano-oscillator
(electron charge e is negative).}
\end{figure}

\begin{figure}[ht]
\includegraphics[width=8.0cm]{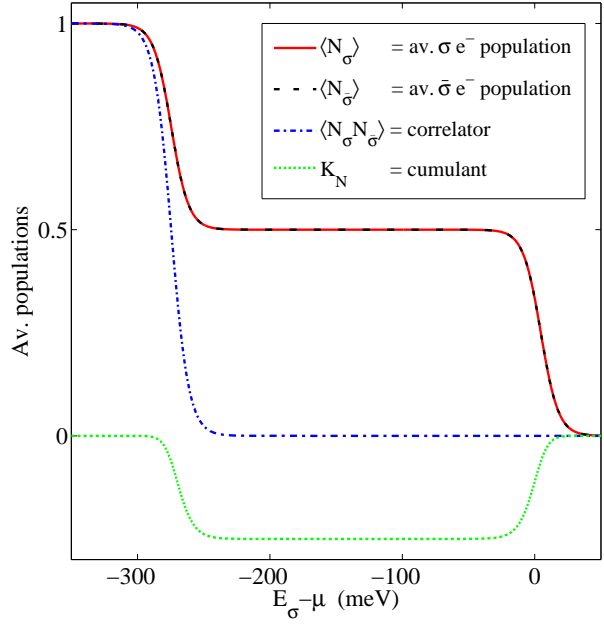}
\caption{\label{fig2} (Color online) Electron populations (for spin
up, $\sigma=1$, in red, for spin down, $\sigma =2$, in dashed
black), populations correlator (dotted-dashed intermediate purple
curve), and cumulant (green dotted curve at the bottom) as functions
of the energy of the nano-oscillator electron state.}
\end{figure}
\begin{figure}[ht]
\includegraphics[width=8.0cm]{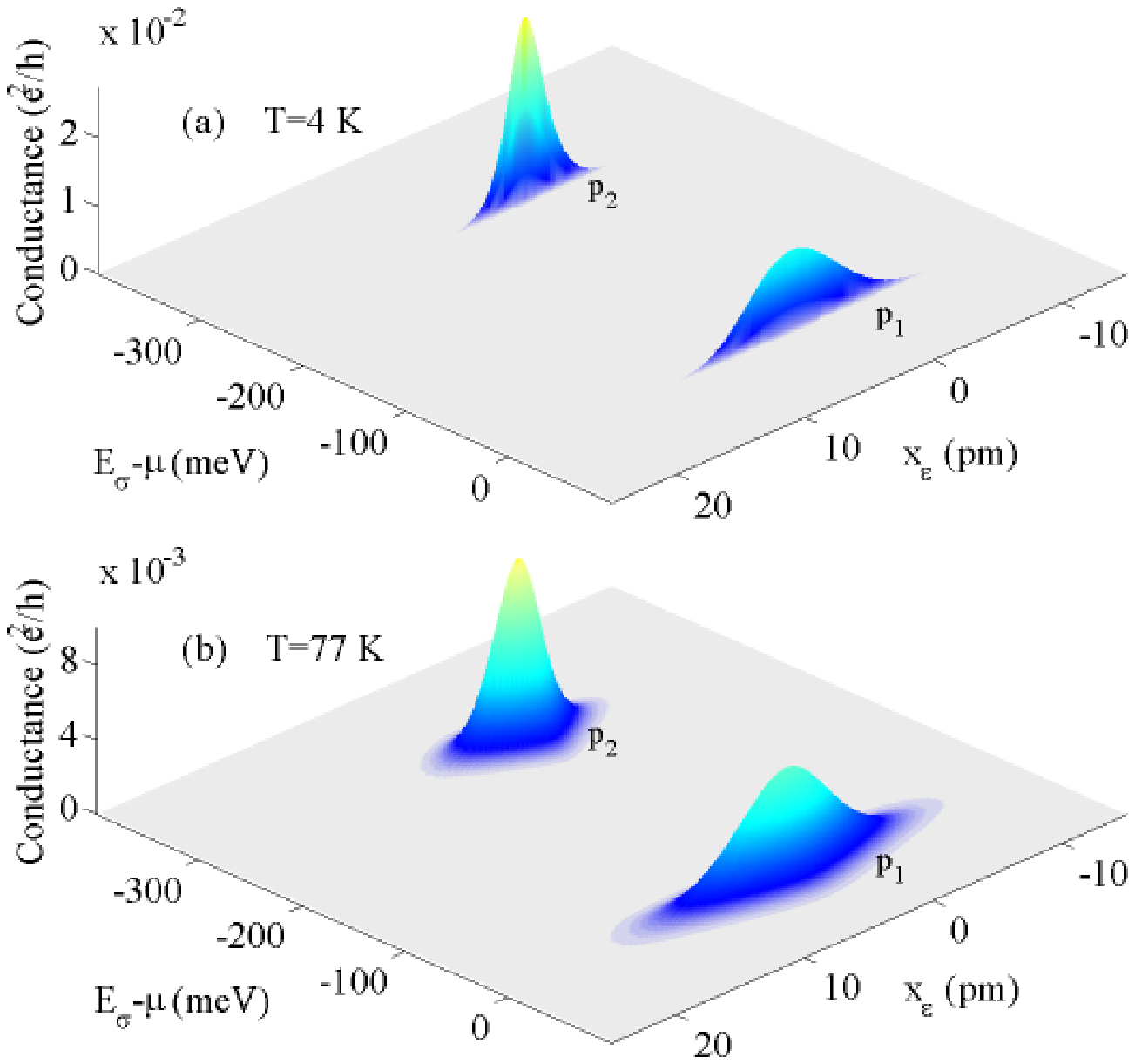}
\includegraphics[width=8.0cm]{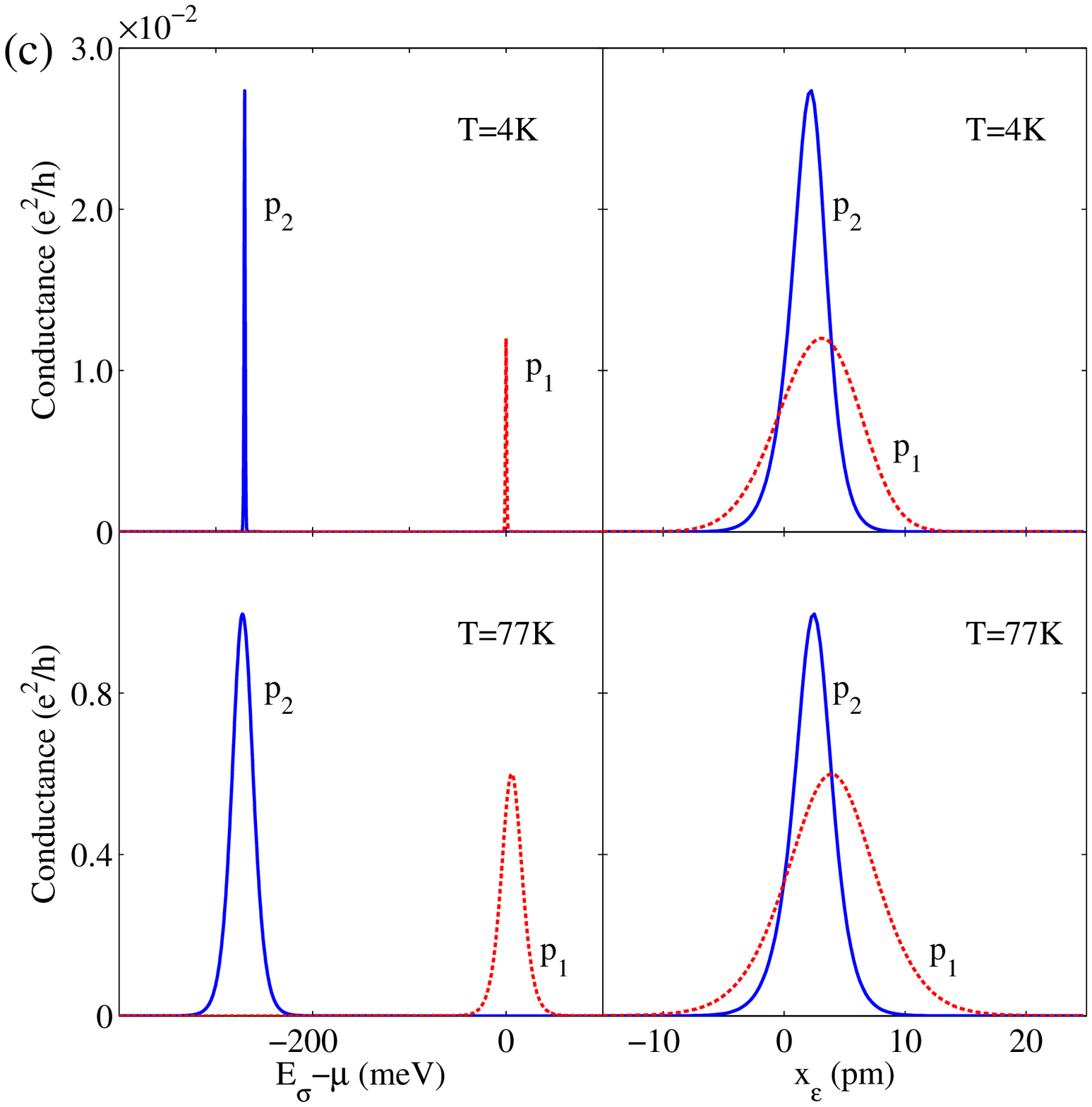}
\caption{\label{fig3} (Color online) Conductance of the
nano-oscillator as a function of the state energy $(E_\sigma-\mu)$
and the oscillator shift $x_{\cal E}$ in an external electric field
${\cal E}$, with a tunneling length $\lambda = 4$ pm, $h\Gamma_L$ =
0.1 meV and $h\Gamma_R$ = 0.002 meV for (a) $T = 4$ K and (b) $T=77$
K. (c) Left: projections to the ``$G$ versus $(E_\sigma-\mu)$''
plane. Right: projections to the ``$G$ versus $x_{\cal E}$'' plane.
The dashed red (continuous blue) peak p$_1$ (p$_2$) denotes the
conductance peak at $E_\sigma=\mu$ ($E_\sigma=\mu-U_0$). When
$x_{\cal E}=0$, p$_1$ and p$_2$ have an equal conductance,
corresponding to the usual Coulumb blockade results.}
\end{figure}
\begin{figure}[ht]
\includegraphics[width=8.0cm]{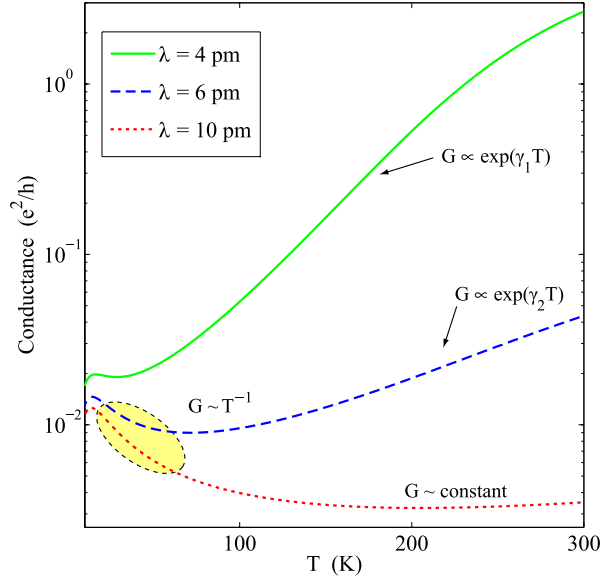}
\caption{\label{fig4} (Color online) Temperature dependence of the
conductance $G(T)$ for various tunneling lengths $\lambda$ and for
an oscillator shift $x_{\cal E}=0$. The yellow (shadowed) ellipse
is the low-temperature regime where $G \sim T^{-1}$.}
\end{figure}
\begin{figure}[ht]
\includegraphics[width=8.0cm]{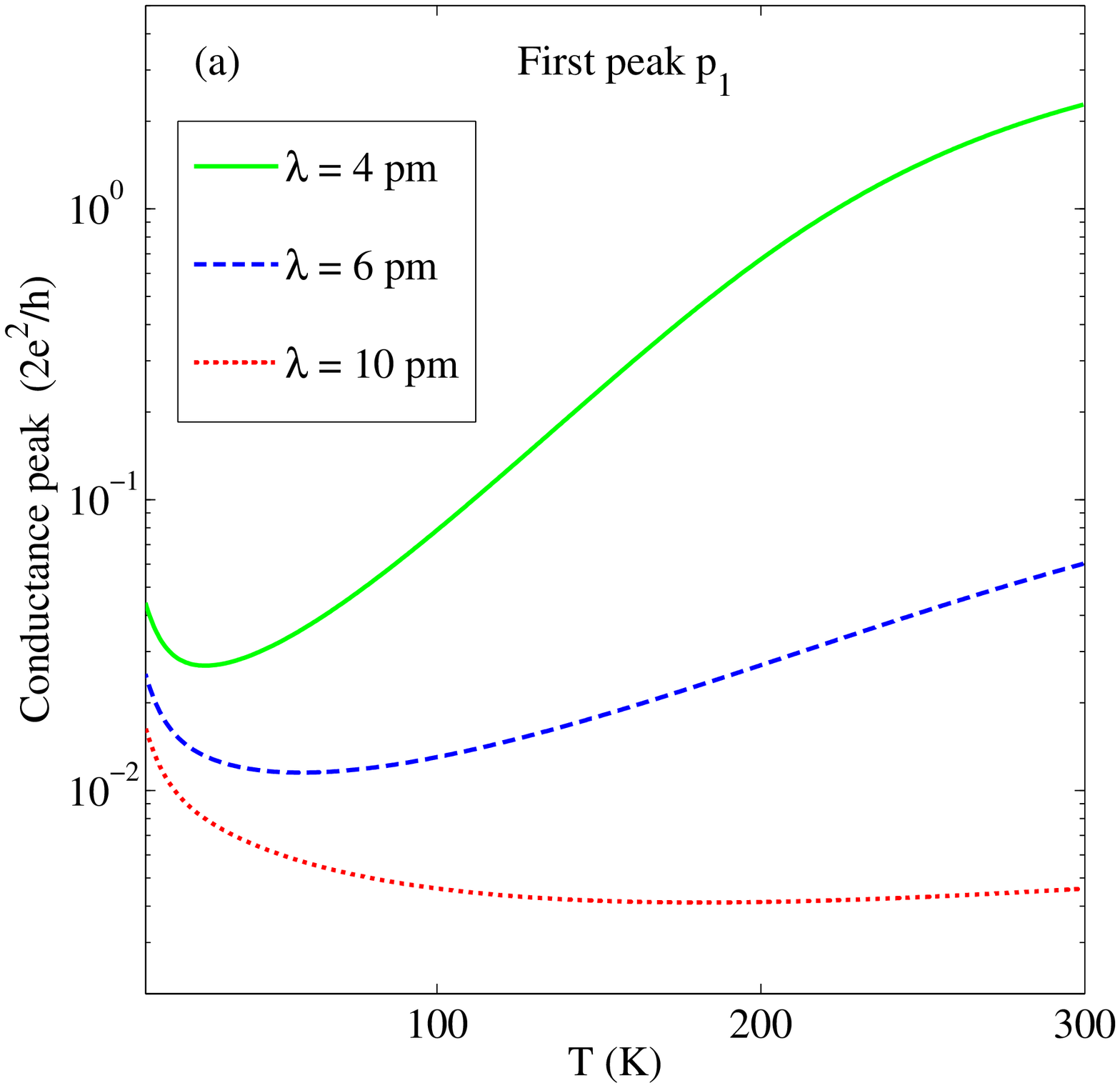}
\includegraphics[width=8.0cm]{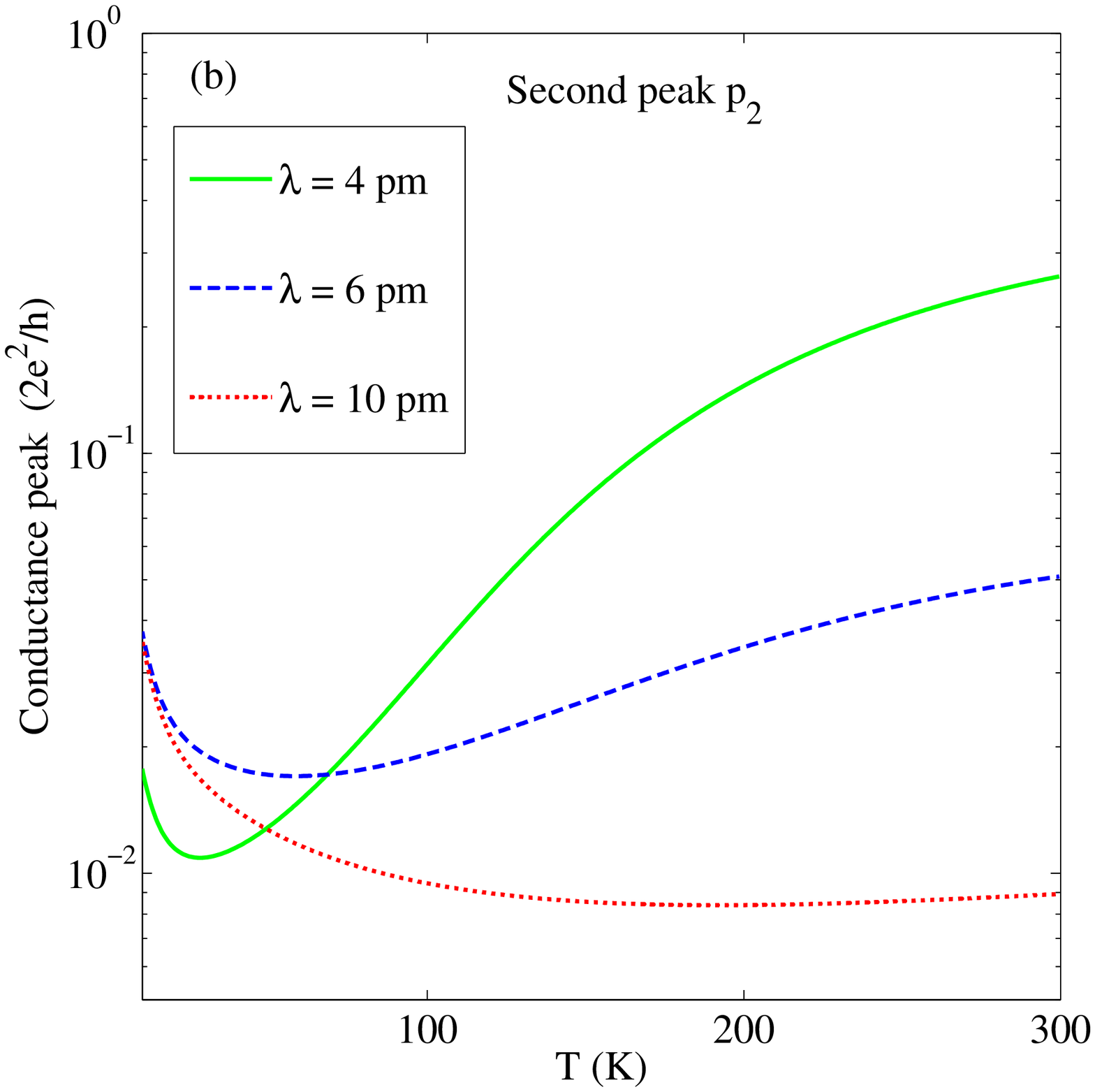}
\caption{\label{fig5} (Color online) Temperature dependence of the
conductance $G(T)$  (for various tunneling lengths $\lambda$ and for
an oscillator shift $x_{\cal E}=5$ pm) for (a) p$_1$ at $E_\sigma =
\mu$ and (b) p$_2$ at $E_\sigma = \mu - U_0$. The conductance can
vary orders of magnitude for small changes in $\lambda$.}
\end{figure}

\end{document}